\documentclass[useAMS,usegraphicx,usenatbib]{mn2e}
\usepackage{epsf,txfonts,subfigure,graphics,graphicx}

\title[Effects of galaxy-halo alignment and adiabatic contraction on 
gravitational lens statistics]
	{Effects of galaxy-halo alignment and adiabatic contraction on gravitational 
	lens statistics}

\author[Q. E. Minor and M. Kaplinghat]
	{Quinn E. Minor$^{1}$, Manoj Kaplinghat$^{1}$ \\
	$^1$Department of Physics and Astronomy, University of California, Irvine, 
	CA 92697.}

\begin{document}

\date{\today}

\pagerange{\pageref{firstpage}--\pageref{lastpage}}\pubyear{2007}

\maketitle

\label{firstpage}

\begin{abstract}
We study the strong gravitational lens statistics of triaxial cold dark matter 
(CDM) halos occupied by central early-type galaxies. We calculate the
image separation distribution for double, cusp and quad
configurations. The ratios of image multiplicities at large
separations are consistent with the triaxial NFW model,  and at small
separations are consistent with the singular isothermal ellipsoid  
(SIE) model. At all separations, the total lensing probability is enhanced by 
adiabatic contraction. If no adiabatic contraction is assumed, naked cusp 
configurations become dominant at $\approx 2.5''$, which is inconsistent with 
the data.  We also show that at small-to-moderate separations ($\lesssim$ 
5'') the image multiplicities depend sensitively on the alignment of
the shapes of the luminous and dark matter projected density profiles.
In constrast to other properties  that affect these ratios, the degree
of alignment does not have a significant effect on the total lensing
probability. These correlations may therefore be  constrained by
comparing the theoretical image separation distribution to a
sufficiently large lens sample from future wide and deep sky surveys
such as Pan-Starrs, LSST and JDEM. Understanding the correlations in the
shapes of galaxies and their dark matter halo is important for future
weak lensing surveys.
\end{abstract}

\section{Introduction}\label{sec:intro}

Statistics of strongly lensed images promise a wealth of information
about luminous and dark matter distributions and the correlations
between them. 
For example, these statistics have yielded general constraints on the 
radial mass profile of early-type galaxies (\cite{maoz93},
\cite{keeton01a},  \cite{wyithe01}, \cite{keeton01},
\cite{rusin-ma01}, \cite{takahashi01},  \cite{sarbu01},
\cite{oguri02}, \cite{oguri02a}, \cite{huterer04},  \cite{kuhlen04}).
Lensing by non-spherical mass distributions shows rich phenomenology
that may be used to probe the distribution of dark matter in galaxies
and clusters. The shape and concentration of the lenses  
Mass distributions that are approximately spherical produce a
higher fraction of double-image systems, whereas more ellipitical
distributions produce more quadruple-image systems. In rare cases,
extreme ellipticities  can produce strongly magnified three-image
configurations known as ``naked  cusp'' systems (or simply
``cusps''). Studying the statistics of different  image multiplicities
(quad, double, and cusp lenses) could help us constrain the shapes of
dark matter halos, the environment of and substructure in these halos
and as we will show the correlations between the shapes of the
luminous and dark matter in galaxies. 

The correlations in the shapes of the galaxy and the dark matter halo
it resides in is important for four reasons. (1) As we show, these
correlations are important to include to make accurate
predictions for the strong lensing probabilities. (2) A large sample
of strongly lensed images will provide a rich data set that can help
us understand galaxy formation better. One of the most interesting
aspects of galaxy formation, and perhaps the least well studied, is
the correlation in the shapes of the luminous and dark matter
components. Lensed image statistics can provide constraints on this
aspect. (3) An important systematic effect for weak lensing surveys is
the spatial correlation in the shapes of the galaxy
shapes, which arise is if galaxy shapes are correlated with their dark
matter halo and halo environment  (\cite{croft00}, \cite{heavens00},
\cite{catelan01}, \cite{crittenden01}, \cite{jing02a},
\cite{mandelbaum06}, \cite{hirata07}). 
All future weak lensing surveys will cover a large fraction of the sky at
great depth and also reveal a great number of strongly lensed quasar
and galaxy images. With an accurate framework in place, one may use
this strong lensing data set to understand systematic effects stemming
from  luminous and dark matter correlations for weak lensing better.
(4) Strong lensing systems are also important for understanding
fundamental physics. As shown by \cite{bolton06a}, these
systems may be used to constrain the post-newtonian gravity parameter
$\gamma$. Constraining this parameter also directly contrains theories
of modified gravity since in this context $\gamma$ is the
ratio of the two scalar potentials in the perturbed FRW metric
(\cite{bertschinger06,zhang07}). 

The distribution of image multiplicities for elliptical lenses has been modeled 
in two regimes: small image separations ($<1''$), in which galaxies become the 
dominant lensing mass, and large (cluster-sized) separations ($>10''$), in 
which dark matter becomes dominant.  \cite{rusin-tegmark01} modeled galaxy 
lenses with a singular isothermal ellipsoid (SIE) mass distribution, comparing 
to lenses from the Cosmic Lens All-Sky Survey (CLASS) and Jodrell-VLA 
Astrometric Survey (JVAS).  They showed that the observed number ratio of 
quads-to-doubles is greatly underestimated by the predictions of the SIE model 
(dubbed the ``quad problem''). While this problem has not been solved 
rigorously, it has been suggested that the effect of massive substructures 
could sufficiently increase the expected quad fraction to be consistent with 
observations (\cite{cohn04}).  Other effects which boost the lens probability 
for quads include misalignment of the centers of galaxies with their host halos 
(\cite{quadri03}) and external shear from lens galaxy environments 
(\cite{huterer05}).

Large separation lenses, on the other hand, were modeled by 
\cite{oguri-keeton04} using the triaxial NFW model based on simulations by 
\cite{jing02}. With an inner slope $\alpha = 1.0$ (pure NFW), the image 
multiplicities are shown to be remarkably different from those of galaxy 
lensing: cusp systems have the highest probability, followed by quads, with 
doubles being least common.  However, because the overall lensing probabilities 
are orders of magnitude smaller than that of galaxy lensing, such dark-matter 
dominated lenses are relatively rare---the only observed lens in this regime is 
a quad at $14.62''$, found in the Sloan Digital Sky Survey (\cite{inada03}). 

In contrast to the paucity of confirmed large-separation lenses, a significant 
fraction of observed gravitational lenses lie in the intermediate region 
between these two extremes, with image separations roughly between $1''$ and 
$10''$.  The total lensing probability in this region has been studied for the 
case of spherical mass distributions (\cite{oguri06}, \cite{ma03}, 
\cite{kochanek01}), but the distribution of image multiplicities for 
nonspherical lenses has not yet been derived.  We note that none of
the single-component elliptical models  mentioned thus far can be
applied in this regime, since both the dark matter  and the central
galaxy contribute significantly at these separations.   Furthermore,
any lensing model including baryons and dark matter must take into  
account the modification of dark matter halos due to baryon cooling, which 
increases the concentration of the inner parts of halos (\cite{kochanek01}, 
\cite{keeton01}).  Although a multi-component model was employed by 
\cite{moeller07} using galaxies with ellipticity, the dark matter was modeled 
by spherical NFW halos, ignoring adiabatic contraction and the triaxiality of 
dark matter halos.

In this paper we derive a distribution of image multiplicities 
that will be approximately valid for all observed image separations, including 
the intermediate region. However, to accurately model this region, we must 
combine the triaxial NFW model, adiabatic contraction, and a central lens 
galaxy. The lensing properties then become sensitive to the ellipticities, 
concentrations, and orientations of both galaxy and halo, in addition to the 
fraction of the total mass contained in the galaxy. This requires a more 
complicated calculation of lensing probabilities than those in the
literature. Despite the complexity of the models considered in this
work, several important factors (considered prevoiously in isolation)
have to be neglected. In particular, we highlight the absence of
dark matter substructure in our model and the effect of the lens
galaxy environment. We will later discuss the expected impact of these
factors. 

We find several interesting features that suggest new methods
of constraining lens mass distributions. For example, the  
ratio of image multiplicities is found to be sensitive to the alignment of the 
major axes of galaxies with that of their host halos.  Further, the transition 
from small- to large-separation lenses is marked by a ``crossing'' of lens 
probabilities when the expected lensing rate of quads overtakes that of 
doubles, and likewise for cusps.  Understanding the precise location of these 
turnover points may provide further means of constraining galaxy-halo mass 
distributions once the sample of observed lens systems becomes sufficiently 
large.

The paper is organized as follows. In \S \ref{sec:imgdist} we will
present a general form for the image separation distribution, and
discuss the relevant model parameters. Section \ref{sec:lensmodel}
will describe the lens model we used and outline   our approach toward 
applying the adiabatic contraction model to triaxial halos.   
Section \ref{sec:calc} covers the details of the calculation, and
results are  given in \S \ref{sec:results}.  Finally we summarize and
discuss our results in  \S \ref{sec:discussion}.

\section{Distribution of Image Separations}\label{sec:imgdist}

The differential image separation distribution is usually defined as the 
probability that a source object at redshift $z_s$ will be multiply imaged with 
an image separation $\Theta$. This is written as (\cite{turner84}, 
\cite{schneider06}),
\begin{equation}
{dP \over d\Theta} = \int dz (1+z)^3 {cdt\over dz} D_{ang}^2(z)
\int d\chi_i p(\chi_i) B\sigma(\Theta,\chi_i,z) {dn\over dM}{dM\over 
d\Theta}\Bigg|_{M(\Theta,\chi_i,z)}\,.
\end{equation}
Here $B\sigma$ is the biased lensing cross section (explained in 
\S\ref{sec:calc}) in units of solid angle, and $n$ is the comoving number 
density of halos.  The lensing optical depth is given by the integral of this 
distribution over all image separations; in this paper we will simply refer to 
it as the total lens probability, though always with respect to a given source 
redshift $z_s$. The $\chi_i$ are the parameters relevant to lensing, whose 
distributions $p(\chi_i)$ are integrated over, and which depend on the 
particular lens model chosen. In this study we take our model to be an 
adiabatically-contracted NFW dark matter halo with a central early-type galaxy 
represented by a Hernquist profile.  The lensing parameters are then

\begin{equation}
\mathbf{\chi} = (c_{e},a/c,b/c,\theta,\phi,f_b,x_b,q_H,\theta_q)\,,
\label{eq:lens_parameters}
\end{equation}
where $c_{e}$ is the triaxial concentration parameter (defined below) of the 
original dark matter halo before adiabatic contraction, $a/c, b/c$ are the axis 
ratios of the halo, ($\theta,\phi$) are the orientation angles of the halo, 
$f_b$ is the baryon mass fraction, $x_b = r_b / r_s$ gives the scale radius of 
the Hernquist profile, $q_H$ is the projected axis ratio of the Hernquist 
profile, and $\theta_q$ is the angle between the projected major axis of the 
Hernquist profile and the projected major axis of the dark matter halo.

The total mass of the lens was not included among the parameters listed above  
because for a given image separation $\Theta$, lens redshift $z$ and set of 
lensing parameters $\chi_i$, the mass is uniquely fixed.

\section{The Lens Model}\label{sec:lensmodel}

In this section we describe the model for the lenses (elliptical
galaxies and clusters). We split the discussion into four parts
dealing with the dark matter halo, the baryonic component, effect of
baryon cooling on the dark matter halo and finally correlations
between the luminous and dark matter mass distributions. 

\subsection{Triaxial dark matter halo}

Throughout this paper we assume a concordance cosmology with $\Omega_M = 0.3$, 
$\Omega_\Lambda = 0.7$, $h = 0.71$, and $\sigma_8 = 0.9$ and use the 
Seth-Tormen  (\cite{sheth02}) mass function of dark matter halos.

We assume the density profile of a triaxial dark matter halo to be 
(\cite{jing02}),
\begin{equation}
\rho(R) = {\delta_{ce}\rho_{crit}(z) \over (R/R_0)(1+R/R_0)^2}\,,
\end{equation}
where
\begin{equation}
R^2 = c^2\left({x^2\over a^2} + {y^2\over b^2} + {z^2\over c^2}\right),~ ~(a 
\leq b \leq c).
\end{equation}

The triaxial concentration parameter is defined in \cite{jing02} as $c_e = 
R_e/R_0$, where $R_e$ is defined so that the mean density within an ellipsoid 
of major axis radius $R_e$ is $\Delta_e \Omega_M(z)\rho_{crit}(z)$ with 
$\Delta_e = 5\Delta_{vir}(c^2/ab)^{0.75}$. The characteristic density is then 
written in terms of the concentration parameter as
\begin{equation}
\delta_{ce} = 
{
\Delta_e \Omega_M(z) c_e^3 
\over 
3 \left[
\ln(1+c_e) - c_e/(1+c_e)
\right]
}\,.
\label{delta_eq}
\end{equation}

We use fitting functions from \cite{jing02} for the distribution of axis ratios 
and concentrations (nearly identical distributions for the axis ratios were 
found in simulations by \cite{allgood06}). The angular orientations are of 
course entirely random ($p(\theta) = (\sin\theta) / 2$, $p(\phi) = 1/2\pi$). 
The fitting function for the median concentration parameter in \cite{jing02} 
has a dependence on the axis ratio $a/c$; however the scatter around their fit 
is quite large, and in practice this only affects the image separation 
distribution significantly at large separations $(\Theta \gtrsim 30'')$.  
Therefore we will neglect the $a/c$ dependence in this paper. 

\subsection{Baryonic component}

For modeling the projected surface density of the galaxy we use an elliptical 
Hernquist profile defined by
\begin{equation}
\Sigma = \Sigma_H(\xi), ~ ~ ~ \xi^2 = q_H x^2 + {y^2\over q_H}\,,
\end{equation}
where $\Sigma_H(r)$ is the projected circular Hernquist profile (see 
\cite{keeton01b} for the analytical formula) and $q_H$ is the projected axis 
ratio.  The ellipse parameter $\xi$ is normalized so that the mass enclosed 
inside a contour of constant surface density is independent of $q_H$; this 
ensures that, when averaged over all angles, the circular profile $\Sigma_H(r)$ 
is recovered. This method of normalization is only approximately correct, since 
the normalization changes if there is a tendency toward oblate or prolate 
galaxies (\cite{chae03}). To obtain an average projected profile of axis ratio 
$q$, one must average over the expected distribution of axis ratios and viewing 
angles that produce the projected axis ratio $q$.  However, our approximation 
is good if we assume equal numbers of oblate and prolate galaxies 
(\cite{chae03}; \cite{huterer05}).

In order to relate the baryon fraction and scale radius to the mass and 
concentration of the host halos, we adopt the mass-luminosity relation proposed 
by \cite{vale04} in terms of bJ-band luminosity. We convert luminosities to 
stellar masses by assuming a universal stellar mass-to-light ratio of $\Upsilon 
= 3.0 h_{70} M_\odot / L_\odot$. We use the scaling relation between g-band 
luminosity and effective radius $R_0$ derived from 9,000 early-type galaxies in 
SDSS by \cite{bernardi03}:
\begin{equation}
{L\over 10^{10.2} h_{70}^{-2} L_\odot} = \left(R_0\over 10^{0.52} h_{70}^{-2} 
kpc\right)^{1.5}\,.
\end{equation}

For a spherical halo, the effective radius is related to the scale radius of 
the Hernquist profile by $r_b = 0.551 R_0$. However, this relation changes for 
an elliptical profile since $R_0$ refers to the observed circular effective 
radius.  Therefore we use a relation of the form $r_b = 0.551 f(q_H) R_0$, 
where $f(q)$ is determined by numerical integration of $\Sigma_H(\xi)$. As in 
\cite{oguri06}, we neglect the difference between bJ-band luminosity and g-band 
luminosity for simplicity.

\subsection{Baryon cooling}

Adiabatic cooling of baryon leads to a increase in the concentration
of dark matter halos. 
We incorporate the effects of baryon cooling on the dark matter halo by 
adopting a modified adiabatic contraction (AC) model. The model is based on the 
original idea of \cite{blumenthal86}, which assumes spherical symmetry and uses 
conservation of angular momentum to calculate the response of dark matter to 
baryon infall. The modified AC model of \cite{gnedin04} incorporates the 
eccentric orbits of particles, though spherical symmetry of the overall density 
profile is still assumed. From simulations based on this model, they presented 
a series of fitting functions that map the radius of a particle after 
contraction to its radius before contraction. The final density profile as a 
function of radius can then be found by inverting the mapping via 
interpolation. The only caveat here is that the simulations in \cite{gnedin04} 
apply only to spherical NFW halos, whereas we are assuming a triaxial model. To 
describe the density profile independent of the effect of AC on halo shape, we 
apply the modified AC model with the replacement $r \rightarrow R$, where the 
triaxial radius $R$ is defined as above. We expect this to be a good 
approximation as long as the axis ratios are not too small.

In addition to making halos more concentrated, adiabatic contraction is also 
expected to make them more spherical, driving the axis ratios closer to one 
especially in the inner parts of the halo (\cite{dubinski94}). This effect is 
seen in simulations by \cite{kazantzidis04} (see also \cite{kazantzidis06}).  
However, the impact of these results on lensing is not well understood.  The 
image separation and cross sections are particularly sensitive to the density 
profile in the inner parts of the halo.  However, the CDM simulation of 
\cite{jing02} and \cite{allgood06} actually show that, in the absence
of baryons, the axis ratios  tend to decrease slightly toward the
center of the halo.  Whether the inner  
axis ratios of the dark matter will end up greater than, less than, or roughly 
equal to the initial average overall axis ratios is unknown.  Therefore, for 
most of the paper we will assume the distribution of axis ratios after 
contraction are not substantially different from those given in \cite{jing02}.  
Later we will investigate the effect on lensing if the axis ratios are 
increased following adiabatic contraction.

\subsection{Correlations between the luminous and dark matter}

Luminous matter cools and forms stars in the gravitational potential
well of the dark matter and hence it is reasonable to assume that the
shapes of the dark and luminous matter mass distributions should be
correlated. There theory explaining this correlation has not been
worked out. Here we ask how these correlations might affect the
lensing probabilities. In particular, we correlate the ellipticity and
projected major axis of galaxy to that of the dark matter halo it
resides in. Previous studies have suggested a possible high degree of 
alignment for disk galaxies (\cite{bosch02}). However, there is not a
clear consensus. We will study the image separation distribution for
the extreme cases, in  which the ellipticities are completely
correlated vs. uncorrelated, and  completely aligned vs. misaligned.
We note that the case where the there is no correlation in the shapes
of the dark and luminous components will lie in between the
two extreme cases we consider.
For the case of completely uncorrelated ellipticities, for simplicity,
we will fix $q_H$, the ellipticity of the galaxy (Hernquist profile),
to a particular value and integrate over the distribution of dark matter
halo ellipticities. When making predictions for a given survey, our
results have to be generalized by averaging over $q_H$ using the
observed distribution. 

\section{Calculating the Image Separation Distribution}\label{sec:calc}

To do the lensing calculations, for each lens we define dimensionless 
coordinates in the lens plane $\vec x = {\vec \xi(z,z_s) / \xi_0(z,z_s)}$, 
where $\xi$ has units of distance in the lens plane and $\xi_0(z,z_s)$ is taken 
to be the Einstein radius of the corresponding spherically symmetric lens at a 
redshift $z$ and source redshift $z_s$. (The angular separation $\Theta$ is 
given by $\xi = \Theta D_L(z)$, where $D_L$ is an angular-diameter distance.)  
We also define corresponding coordinates in the source plane, $\vec X = {\vec 
\eta(z,z_s) / \eta_0(z,z_s)}$ where $\eta_0 = \xi_0 {D_s / D_L}$. The cross 
sections and image separations in these units will be denoted by $\tilde\sigma$ 
and $\tilde\Theta$, respectively.

\subsection{Projected mass density profile}

Before solving the lens equation we must first obtain the projected mass 
density profile, $\Sigma(\vec r)$. This is generated numerically by taking the 
triaxial adiabatically-contracted NFW profile mentioned above, assuming some 
orientation angles $\theta$ and $\phi$, and integrating along the line of 
sight. However, this method can be simplified considerably by a judicious 
change-of-variables.  We give a brief summary of the method here; for more 
details we refer the reader to \cite{oguri03} and \cite[pg.~40]{oguri04}, where 
the method is applied to triaxial NFW halos.  The triaxial radius $R$ (defined 
above) is written in terms of the observer's coordinates $(x',y',z')$, with 
$z'$ axis along the line of sight, as
\begin{equation}
R = \sqrt{f z'^2 + g(x',y') z' + h(x',y')} = \sqrt{{z'_*}^2 + \zeta^2}\,,
\label{fgh_eq}
\end{equation}
where $f,g,h$ are all functions of the axis ratios and orientation angles, but 
$f$ has no dependence on $x',y'$. 
Further $z'_* \equiv \sqrt{f}(z' + g/2f)$, $\zeta^2 \equiv h - g^2/4f$
and with this change of variables, we have 
\begin{equation}
\Sigma(\zeta) = \int_{-\infty}^{\infty}\rho(R)dz' = {1\over \sqrt{f}} 
\int_{-\infty}^{\infty}\rho\left(\sqrt{ {z'_*}^2 + \zeta^2}\right)dz'_*
\end{equation}

The $\zeta$ parameter can be shown to correspond to the projected isodensity 
curves, the two-dimensional analogue of $R$ defined above. It can be written as
\begin{equation}
\zeta^2 = {x'^2\over q_x^2} + {y'^2\over q_y^2},
\end{equation}
where the x-axis is taken to lie along the projected major axis of the 
isodensity curve, and $q_x$ and $q_y$ are complicated functions of the axis 
ratios and orientation angles.  We can see from the equations above that the 
result of changing the axis ratios and orientations is to scale the projected 
density by a factor $F_z = 1/\sqrt{f}$, and scale the x- and y-coordinates by 
the factors $q_x$ and $q_y$ respectively.  Therefore, we need only calculate 
the above integral for the spherically-symmetric case, and then apply these 
scalings to obtain the projected density profile for any combination of axis 
ratios and orientation angles.  This vastly simplifies the calculation.  
Finally, we superimpose this with the Hernquist profile described above, and 
divide by the critical lensing density $\Sigma_{crit}$ to arrive at a 
dimensionless ``kappa'' profile $\kappa(\vec x)$. 

\subsection{Lensing cross-sections \label{sec:cross}}

After generating the kappa profile, we then calculate the specific cross 
sections for lensed sources of different image multiplicities, depending on 
whether they are double, quad, or naked cusp image configurations.  The cross 
sections and image separations are calculated by a Monte Carlo method where a 
thousand sources are placed randomly within the caustics of the gravitational 
lens. For each source point we solve the lens equation numerically, using the 
integration formulas of \cite{schramm90} for the deflection and magnification 
of elliptical mass profiles to obtain the corresponding images and 
magnifications (\cite{keeton01b}). To find the images we use an adaptive radial 
grid where the grid cells recursively divide themselves several times, with two 
more subdivisions when a cell is in the vicinity of a critical curve.  Images 
are then found by mapping each trapezoidal cell to the source plane and testing 
whether the source point falls inside the mapped region.  (Each cell is 
actually split into triangles first to avoid the ambiguity of having
convex vs. concave cells; see \cite{bartelmann03} for details.) 

We exclude all lens systems with a flux ratio between the brightest two images 
greater than 15:1, in accord with the flux ratio selection function of radio 
searches using the VLA. In practice, such large flux ratios are common only in 
the isothermal case with double images. For the adiabatically contracted model, 
less than 0.1\% of lensing events result in such large flux ratios, whereas the 
fraction for an isothermal lens is roughly 15-20\%. 

The image separations, defined here as the maximum separation between any two 
pairs of images in an image configuration, are then averaged over all sources 
to obtain the average image separation $\tilde\Theta$. The cross sections can 
be obtained by taking the fraction of sources which produce a given image 
configuration (quad, double, or cusp), multiplied by the total cross-section.  
In addition, if we also weigh the sources by their magnifications 
appropriately, we can compute the magnification bias.  Assuming the sources 
follow a power-law luminosity function $\phi_L(L) \propto L^{-\beta}$ 
(generally true for luminosities relevant to lens searches), the biased 
cross-section is written as
\begin{equation}
B\tilde\sigma = \int d^2X {\phi_L(L/\mu)/\mu\over \phi_L(L)} = \int d^2X 
\mu^{\beta-1}\,,
\end{equation}
where $\mu$ is the magnification, and we are integrating over the 
multiply-imaged region of the source plane in dimensionless units.

There is, however, an ambiguity in this formula: should the lens systems be 
weighted by the total magnification of all the images, or by one of the 
brighter images?  The answer depends on the particular survey and the method of 
selecting lens candidates.  For most surveys to date, such as JVAS/CLASS, the 
number of observed quasars is sufficiently small that each quasar can be 
returned to and observed in great detail. In such cases, for small-to-medium 
image separations the brightness of the lens system is seen as the total 
brightness of all the images, so the total magnification should be used in the 
bias (\cite{takahashi01}, \cite{cen94}). However, for large angular 
separations, a lens system will only be identified if a second image is 
independently resolved in the survey, and therefore the magnification of the 
second-brighest image should be used.

Looking ahead to future large-scale synoptic surveys using telescopes such as 
LSST, the number of observed lenses may be orders of magnitude larger and each 
lensed object may not be observed in such detail. Lensed quasars, for example, 
can be recognized using image subtraction methods that obviate the need for 
extensive follow-up observations (\cite{kochanek06}). In that case the 
second-brightest image bias may be more appropriate as the more conservative 
bias. This will almost certainly be the case when searching for galaxy lenses 
in large surveys such as the SuperNova Acceleration Probe (SNAP).  We will 
consider both types of bias in this study.

From the method above we can obtain the biased cross sections for each image 
configuration type separately as well as the total biased cross section.  These 
will be written as $\tilde\sigma_i$ where $i=2,3,4,0$ for doubles, cusps, 
quads, and the total cross section respectively. Throughout this paper we will 
assume a power-law index $\beta = 2.1$ for the luminosity function, which is 
consistent with flat-spectrum radio sources seen by surveys such as CLASS 
(\cite{myers03}, \cite{rusin-tegmark01}).

The principle difficulty in doing the above calculation is that the cross 
sections have a complicated dependence on almost \emph{all} the lensing 
parameters $\chi_i$, in addition to the redshift $z$.  However, a few 
simplifications can be made.  We can reduce the number of parameters by one if 
we transform the orientation and ellipticity parameters to their projected 
counterparts:
\begin{equation}
(a/c,b/c,\theta,\phi) \longrightarrow (q_x,q,F_z)
\end{equation}
Here $q_x$ and $F_z$ are the same parameters introduced earlier, and $q = q_y / 
q_x$ is the projected axis ratio. The parameter $q_x$ is the ratio of the 
projected major axis to the triaxial major axis of the halo. It might be 
objected that we cannot perform this reduction since the characteristic density 
$\delta_{ce}$ has an explicit dependence on the axis ratios $a/c, b/c$ through 
the triaxial overdensity $\Delta_e$ (see eq.$~\ref{delta_eq}$). However, this 
dependence simply scales the kappa profile and thus can be accounted for by 
redefining the $F_z$ parameter as $F_z = (c^2/ab)^{0.75}/ \sqrt{f}$, 
where $f$ is defined in eq.~\ref{fgh_eq}.

For simplicity we neglect scatter in the mass-luminosity relation and halo 
concentration. We will investigate later (see \S \ref{sec:results}) how the 
scatter might affect the calculation.

\subsection{Image separations}

Another simplification can be made by first noting that the average image 
separation, in the absence of ellipticity, is very well approximated by $\Theta 
\approx 2 R_E$ where $R_E$ is the Einstein radius; or, in the units defined 
above, $\tilde\Theta \approx 2$.  Introducing ellipticity changes the 
proportionality by a certain amount; however, $c_e$, $q_x$, $F_z$ and $z$ 
simply scale the Einstein radius without significantly altering this relation.  
We find that the image separations resulting from the dark matter halo is fit 
well by the form,
\begin{equation}
\tilde\Theta_{dm}(q) = \alpha(M) - \beta(M)\left[exp[(1-q)\gamma(M)] - 
1\right]\,,
\end{equation}
where $M$ is the mass of the halo. Likewise, the Hernquist component is well 
fit by
\begin{equation}
\tilde\Theta_H(q_H) = \alpha(M)\left[0.5(1+q_H^{\beta(M)})\right]^{\gamma(M)}\,.
\end{equation}
Both models are conceived so that $\alpha(M) = 2$ when $q = 1$.  Finally, we 
model the total image separation $\tilde\Theta$ by a fitting function of the 
form
\begin{equation}
\tilde\Theta = \alpha(M) \left[{(R_{E,dm}\tilde\Theta_{dm}(q_{dm}))^{\beta(M)} 
+ (R_{E,H}\tilde\Theta_H(q_H))^{\beta(M)} \over 
(R_{E,dm}\tilde\Theta_{dm}(q_{dm}=1))^{\beta(M)} + 
(R_{E,H}\tilde\Theta_H(q_H=1))^{\beta(M)}}\right]^{\gamma(M)\over\beta(M)}\,.
\end{equation}

The advantage of this approach is that, for a given image separation $\Theta$, 
we can easily find the total mass of the lens as a function of the other 
parameters, $M(c,q,q_H,q_x,F_z)$, and likewise for $dM/d\Theta$. This 
reduces the number of parameters by one. Thus we are left with the following 
independent variables on which the cross sections depend:
\begin{equation}
\tilde\sigma_i \rightarrow \tilde\sigma_i(q,q_H,q_x,F_z,z)\,.
\end{equation}

\subsection{Projected ellipticities}

To simplify the analysis, rather than taking $q_H$ as an independent variable 
we will investigate two cases: first, we assume the ellipticities of galaxy and 
halo are strictly correlated; second, we assume different fixed values of 
$q_H$.  Our motivation is that since the ellipticities of the 
adiabatically-contracted halos are still not well understood, the extent to 
which the ellipticities are correlated is unknown.  Undoubtedly the true 
distribution of ellipticities will lie somewhere between these two extreme 
cases. If we assume that $q$ and $q_H$ are completely uncorrelated, we can 
simplify the analysis by fixing $q_H$ at various values including its mean 
observed value. Several studies have analyzed the isophotal shapes of E/S0 
galaxies in various surveys, the distribution of which is typically well fit by 
a Gaussian: $dn / d\epsilon \propto exp[-(\epsilon-\epsilon_0)^2/2\Delta 
\epsilon^2]$, where $\epsilon = (1-q_H^2)/(1+q_H^2).$ \cite{keeton97} studied 
galaxies in the Coma cluster and found $\epsilon_0 = 0.26, \Delta \epsilon = 
0.33$, whereas \cite{hao06} studied 847 E/S0 galaxies in SDSS and found
$\epsilon_0 = 0.22, \Delta \epsilon = 0.14$. The Coma distribution has a mean 
value at $\langle q_H \rangle = 0.67$, which we will use for the uncorrelated 
case unless otherwise noted. On the other hand, if we make the simplifying 
assumption that  $q$ and $q_H$ are strictly correlated, we can relate them 
directly by $q_H = q + \Delta q$ where $\Delta q$ is a fixed increment (or 
simply $q_H = 1.0$ if $q > 1.0 - \Delta q$). We will investigate both the 
correlated and uncorrelated cases, taking $\Delta q = 0.1$ for the correlated 
case.

We will calculate $dP /d\Theta$ at specific values of $z$; these 
variables will be integrated over at the end. To do the integral we first 
calculate the values of $\sigma$ over a regular grid in the remaining three 
parameters $(q,q_x,F_z)$ and then interpolate to find its value at any given 
point.  The interpolation is done by an N-dimensional analogue of bicubic 
interpolation, whereby the function values, its partial derivatives, and 
higher-order mixed partial derivatives are tabulated on a regular grid (the 
derivatives being calculated in this case by finite differencing across 
adjacent grid points).  The interpolating function is a cubic polynomial in the 
N variables whose coefficients are uniquely determined by matching its values 
and derivatives to the tabulated values on the grid. Thus the resulting 
function is constrained to have the tabulated values on the grid, and also to 
vary smoothly from point to point. (For a detailed derivation of the 
three-dimensional case, see \cite{lekien05}.) With $\sigma$ in hand, we 
integrate over the axis ratios and line-of-sight angles (which are the 
remaining lens parameters in eq.~\ref{eq:lens_parameters}) using the Vegas 
algorithm, and finally integrate over redshift to obtain the image separation 
distribution.

\section{Results}\label{sec:results}

\begin{figure}
	\includegraphics[height=1.0\hsize,angle=-90]{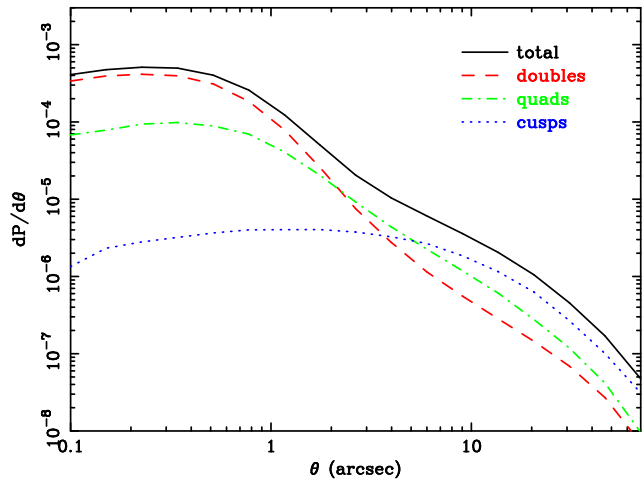}
\caption{{\em Distribution of image separations for double, quad, and cusp
  image configurations.}  
Galaxy and dark matter halo shapes are aligned, with correlated
ellipticities ($q_H =  q + 0.1$).  The ``second brightest image''
magnification bias is used here.  No cut-off due to the finite angular
resolution has been imposed in this or any of the other figures.} 

\label{fig:dpdtheta-aligned-corr}
\end{figure}

Throughout this paper we plot the image separation distribution for a source 
placed at a redshift of $z_s = 2.0$; the results are qualitatively the same for 
other source redshifts.  First we assume the major axes of galaxy and halo are 
aligned, with correlated ellipticities.  Using the ``second brightest image'' 
magnification bias, the results are shown in 
fig.~\ref{fig:dpdtheta-aligned-corr}.  The lensing probability for quads 
overtakes that of doubles at $~2''$, whereas cusps become dominant at $~5''$.  
Note that here we have not taken into account the angular resolution of 
surveys, which cuts off the distribution sharply at small image 
separations.  The lensing probability for CLASS, for example, cuts off at 
$~0.3''$.

\subsection{Effect of magnification bias}

\begin{figure}
	\includegraphics[height=1.0\hsize,angle=-90]{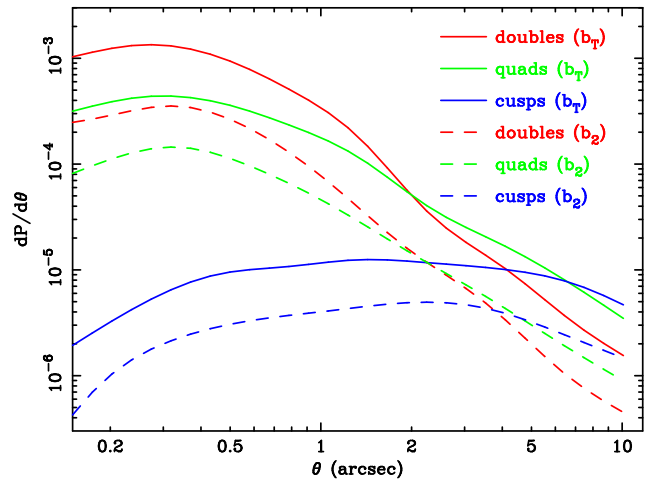}
\caption{
{\em Comparison of the ``total image'' magnification bias (labeled $b_T$)
and the ``2nd-brightest image'' bias (labeled $b_2$).} (See \S
\ref{sec:cross} for details.) We assume a power-law luminosity  
function, $\phi(L) \sim L^{-2.1}$. The shape of all galaxies are
taken to be aligned with their host dark matter halo and we have fixed
galaxy axis ratio to $q_H = 0.67$.} 
\label{fig:dpdtheta-bias}
\end{figure}

We also plot the image multiplicities using the total magnification bias (which 
we will abbreviate as $b_T$.  In fig.~\ref{fig:dpdtheta-bias} we compare to the 
``second-brightest image'' magnification bias ($b_2$). We see that the image 
multiplicities are relatively unchanged, although the lensing probability for 
each image configuration is larger by on overall factor of ~3.5 compared to the 
$b_2$ bias (note, however, that this factor will depend on the slope of the 
luminosity function, which we have taken as $\eta = 2.1$).  Surprisingly, the 
ratio of cusps/doubles and cusps/quads is slightly larger using the $b_2$ bias, 
which is clearly shown by the fact that the turning points, when cusps overtake 
doubles and quads, are shifted to the left.  This is because cusps systems 
nearly always have at least two (and often three) bright images, whereas 
doubles and quads often have one image (or two for quads) that is significantly 
brighter than the others.  So using the total magnification bias actually 
boosts the number of quads and doubles relative to cusps.

Any attempt to solve the ``quad problem'' of radio lenses (described in \S 
\ref{sec:intro}) can then be rephrased as follows: given a cutoff at small 
image separations, does the ``turnover point'' for quads/doubles occur at a 
sufficiently small image separation such that the total number of quads is 
roughly equal to that of doubles? Given the cutoff separation for CLASS, from 
the graph it is obvious that this condition is not satisfied; the majority of 
lenses still occur before $2''$, where doubles are dominant. It seems 
reasonable, however, that the inclusion of spirals, substructure and external 
shear from lens galaxy environments may push the turnover point to sufficiently 
small image separations to solve the quad problem.

\subsection{Comparison to NFW profile and Singular Isothermal
  Ellipsoid}

\begin{figure*}
	\centering
	\subfigure[total]
	{
		\includegraphics[height=0.26\hsize,width=0.28\hsize,angle=-90]{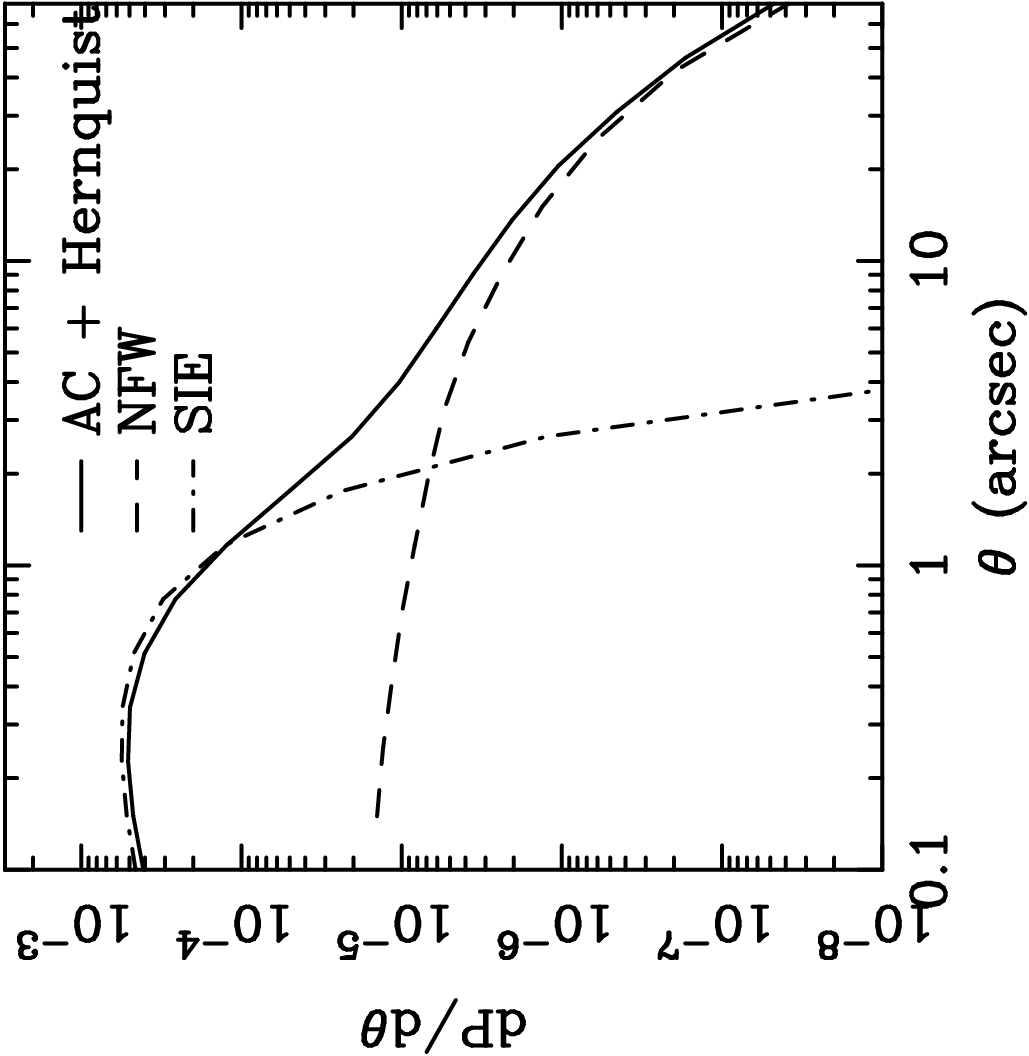}
		\label{fig:dpdtheta-ac-nfw-sie-total}
	}
	\subfigure[doubles]
	{
		
\includegraphics[height=0.23\hsize,width=0.28\hsize,angle=-90]{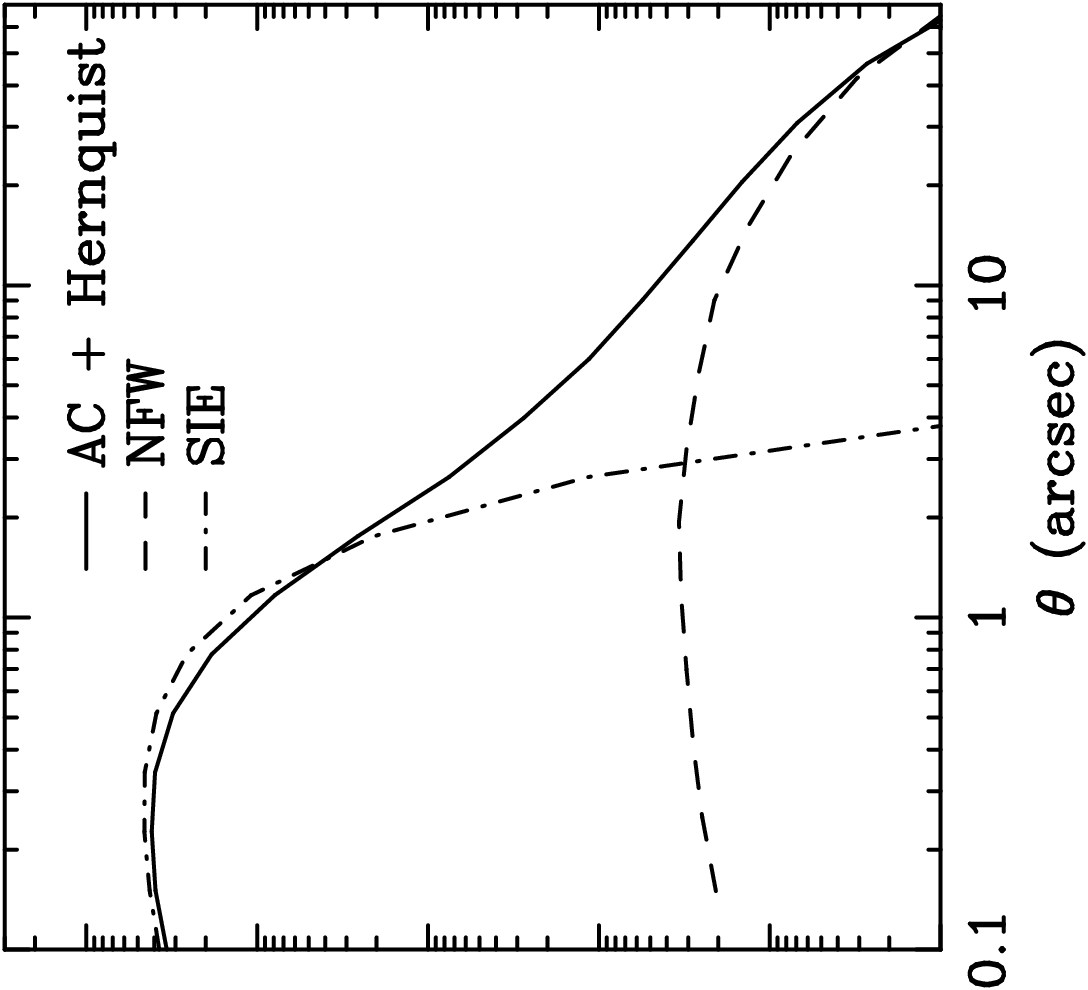}
		\label{fig:dpdtheta-ac-nfw-sie-double}
	}
	\subfigure[quads]
	{
		\includegraphics[height=0.23\hsize,width=0.28\hsize,angle=-90]{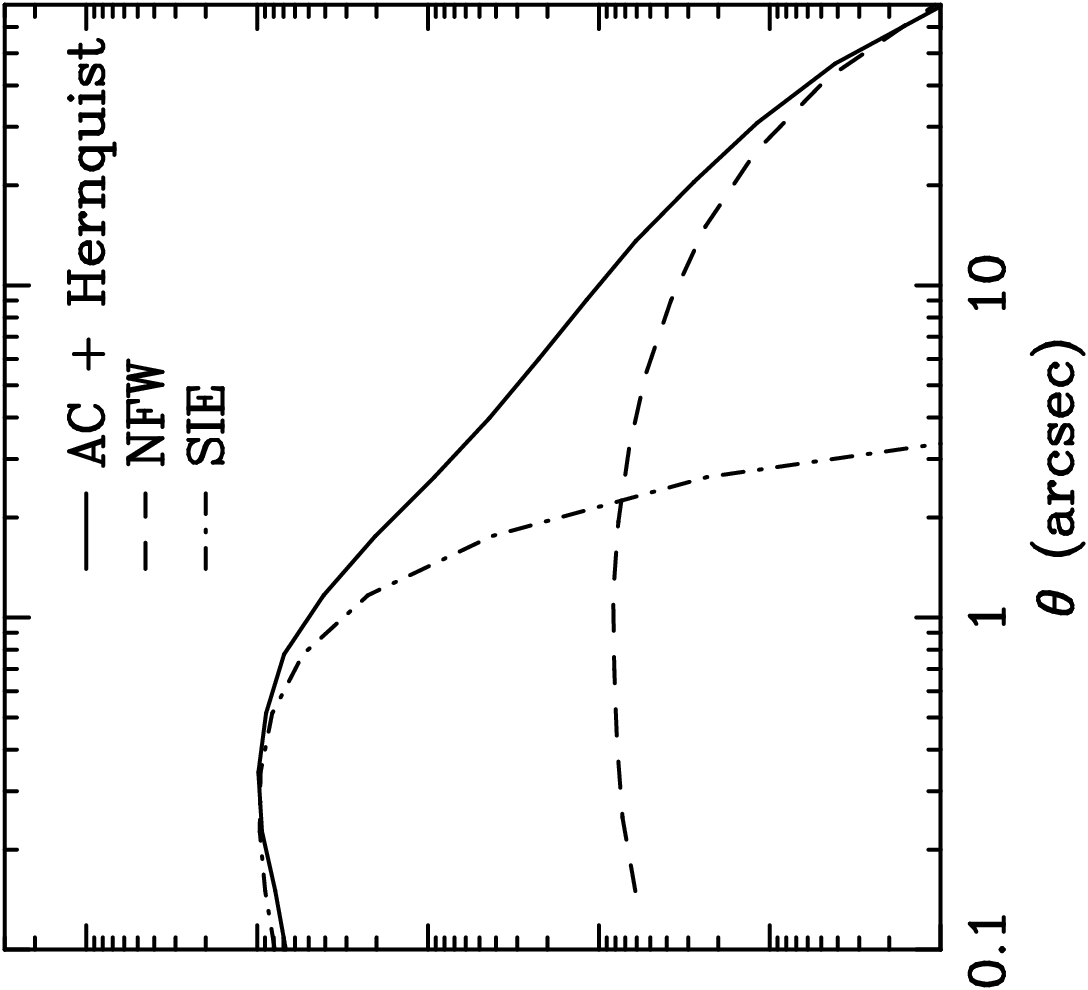}
		\label{fig:dpdtheta-ac-nfw-sie-quad}
	}
	\subfigure[cusps]
	{
		\includegraphics[height=0.23\hsize,width=0.28\hsize,angle=-90]{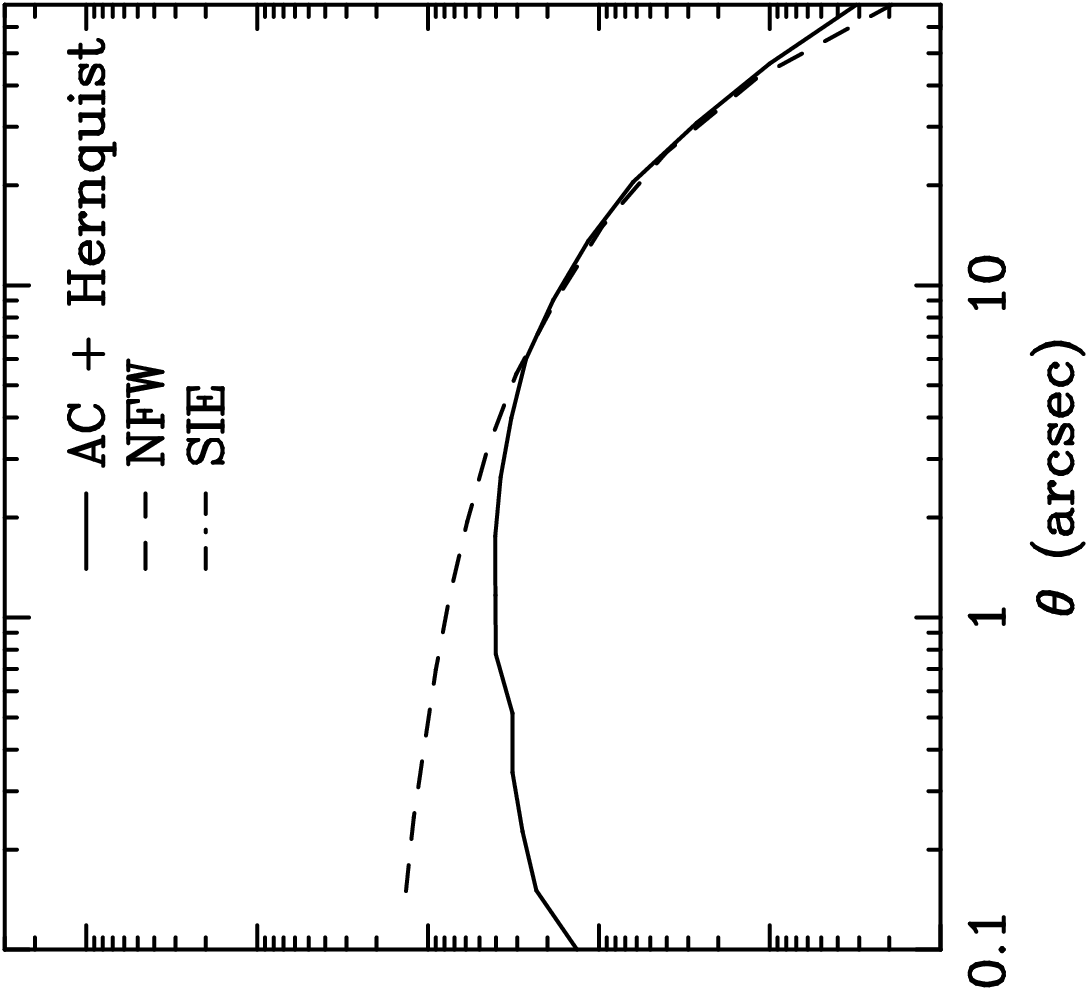}
		\label{fig:dpdtheta-ac-nfw-sie-cusp}
	}
	\caption{{\em Comparison of the full model to distribution of image
          multiplicities for an NFW density profile and Singular
          Isothermal Ellipsoid (SIE) density profile.} Going from left
        to right we have the comparisons for the the total lensing
        probability, doubles, quads, and cusps. The lensing
        probability for cusps in the SIE model is smaller than
        $10^{-8}$ and not shown here.} 
	\label{fig:dpdtheta-ac-nfw-sie}
\end{figure*}

As a consistency check, it is instructive to compare this distribution to that 
of lensing by galaxies represented by an SIE profile, and dark matter halos 
represented by an NFW profile. These cases are plotted in 
fig.~\ref{fig:dpdtheta-ac-nfw-sie}. Interestingly, quads and doubles are 
significantly enhanced compared to NFW even at separations as large as $20''$.

It is somewhat surprising 
that our model agrees so well with SIE at low separations. The agreement can be 
partly understood by comparing the ellipticities of the dark matter and 
galaxies.
For the SIE model we used 
the \cite{hao06} ellipticity distribution, whereas the galaxy ellipticities in 
our AC model were directly correlated with their dark matter halos according to 
$q_{gal} = q + 0.1$.  The $q$-distribution of the dark matter halos is from 
\cite{jing02}, and this distribution depends on the mass of the halo.
The image separation distribution is  
dominated at low separations by $10^{11}M_\odot-10^{12}M_\odot$ systems.  
We verified that the distribution in this range roughly coincides with  
the observed distribution in \cite{hao06}, though the halo ellipticity
distribution has a significantly larger tail at low $q$-values.  Indeed,  
since naked cusps appear only at $q \lesssim 0.4$ for an SIE lens, cusps have 
negligible lensing probability in the SIE model $(dP / d\Theta \lesssim 
10^{-9})$. From the $q$-distribution alone, therefore, one might expect the AC 
model to yield slightly more quads (and less doubles) than the SIE model.

However, ellipticity is not the only factor determining the quad-to-double 
ratio. In fact, when compared at identical ellipticities, our model results in 
a smaller quad-to-double ratio compared to SIE.  The reason is that, while the 
profile of the adiabatically contracted halo is very close to isothermal, the 
addition of a Hernquist profile actually steepens the slope of the density 
profile in the vicinity of the Einstein radius. This enlarges the radial 
caustic and results in more doubles compared to quads and cusps. The net effect 
is that the slightly higher ellipticities in the AC model is nearly balanced 
out by the steeper density profile compared to SIE. In view of these 
considerations, we consider the agreement of our model with SIE (at least for 
doubles and quads) to be somewhat fortuitous.


\subsection{Alignment of projected major axes}

\begin{figure*}
	\centering
	\subfigure[doubles]
	{
		
		\includegraphics[height=0.265\hsize,width=0.28\hsize,angle=-90]{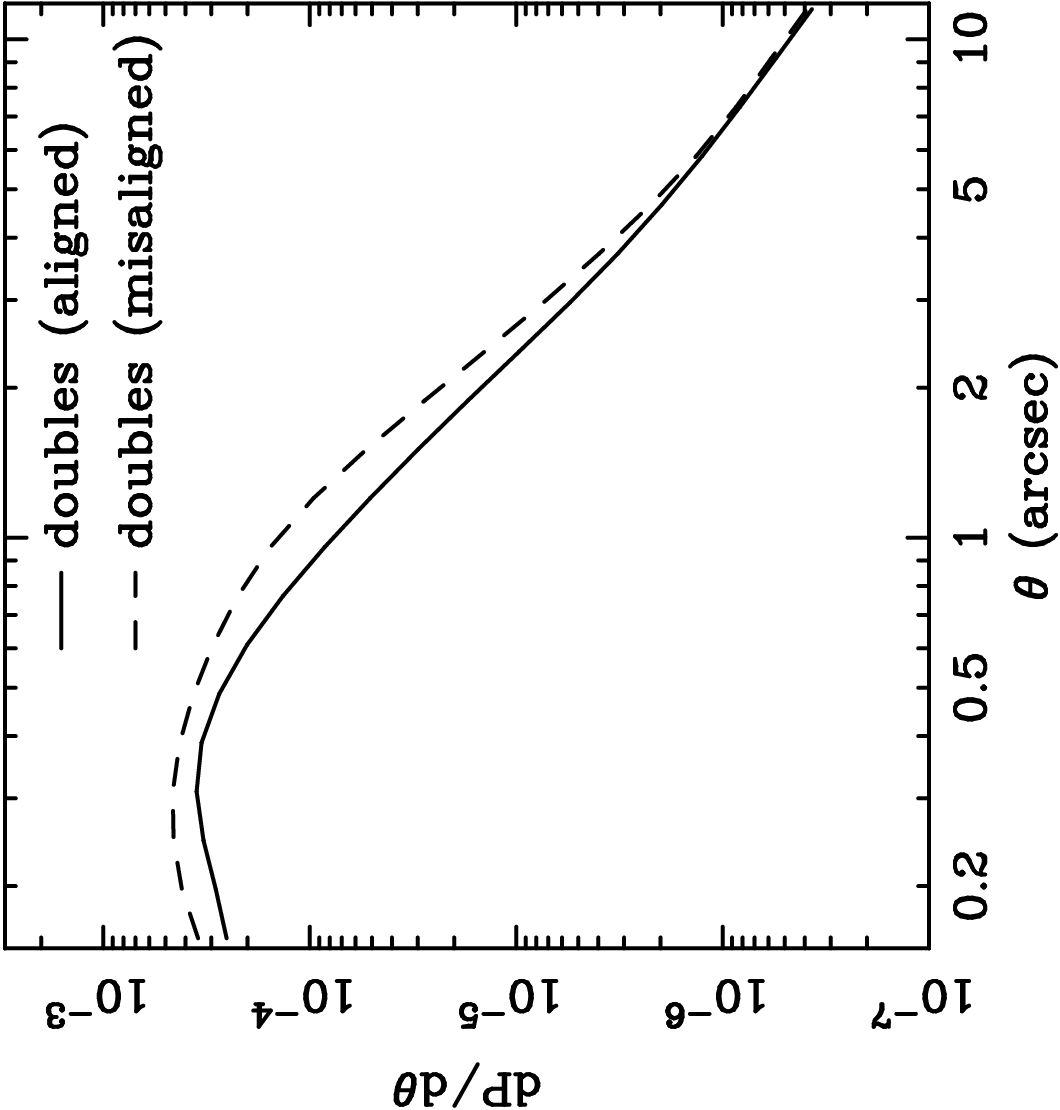}
		\label{fig:dpdtheta-0-90.doubles}
	}
	\subfigure[quads]
	{
		\includegraphics[height=0.23\hsize,width=0.28\hsize,angle=-90]{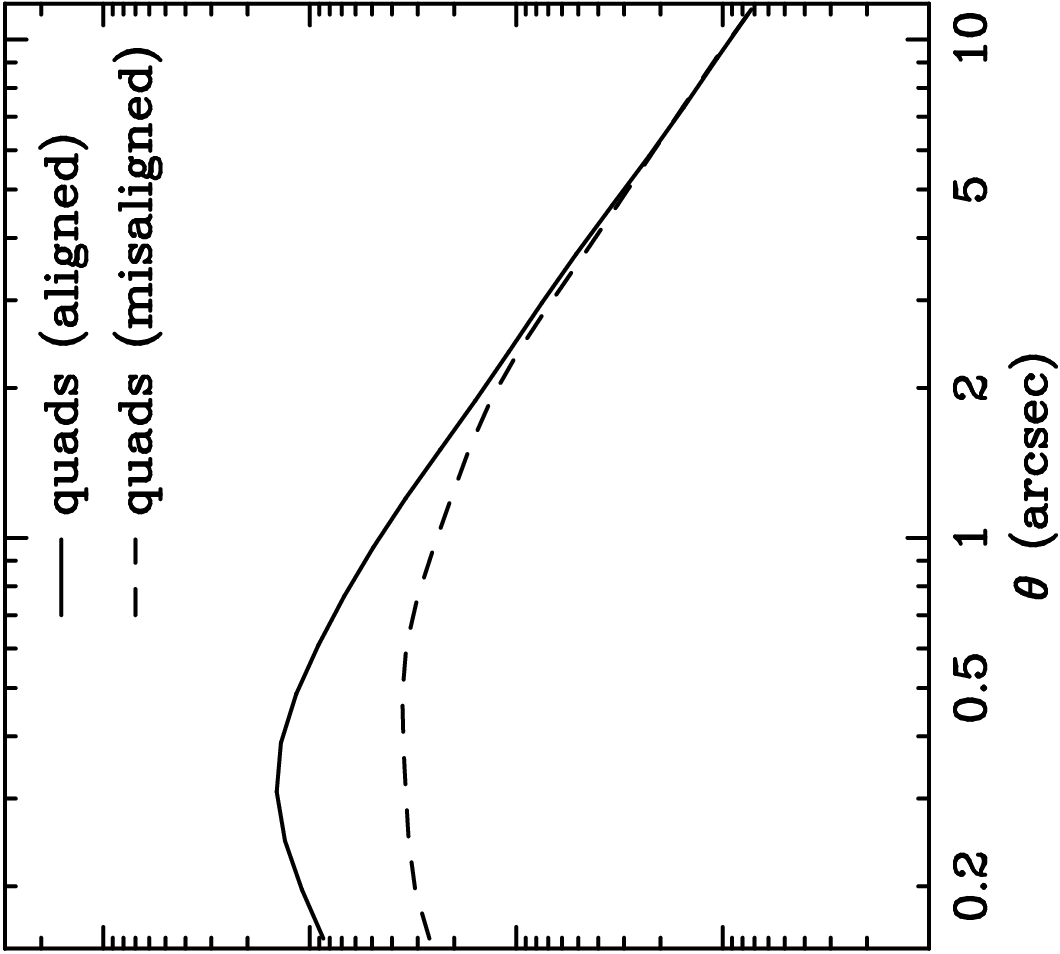}
		\label{fig:dpdtheta-0-90.quads}
	}
	\subfigure[cusps]
	{
		\includegraphics[height=0.23\hsize,width=0.28\hsize,angle=-90]{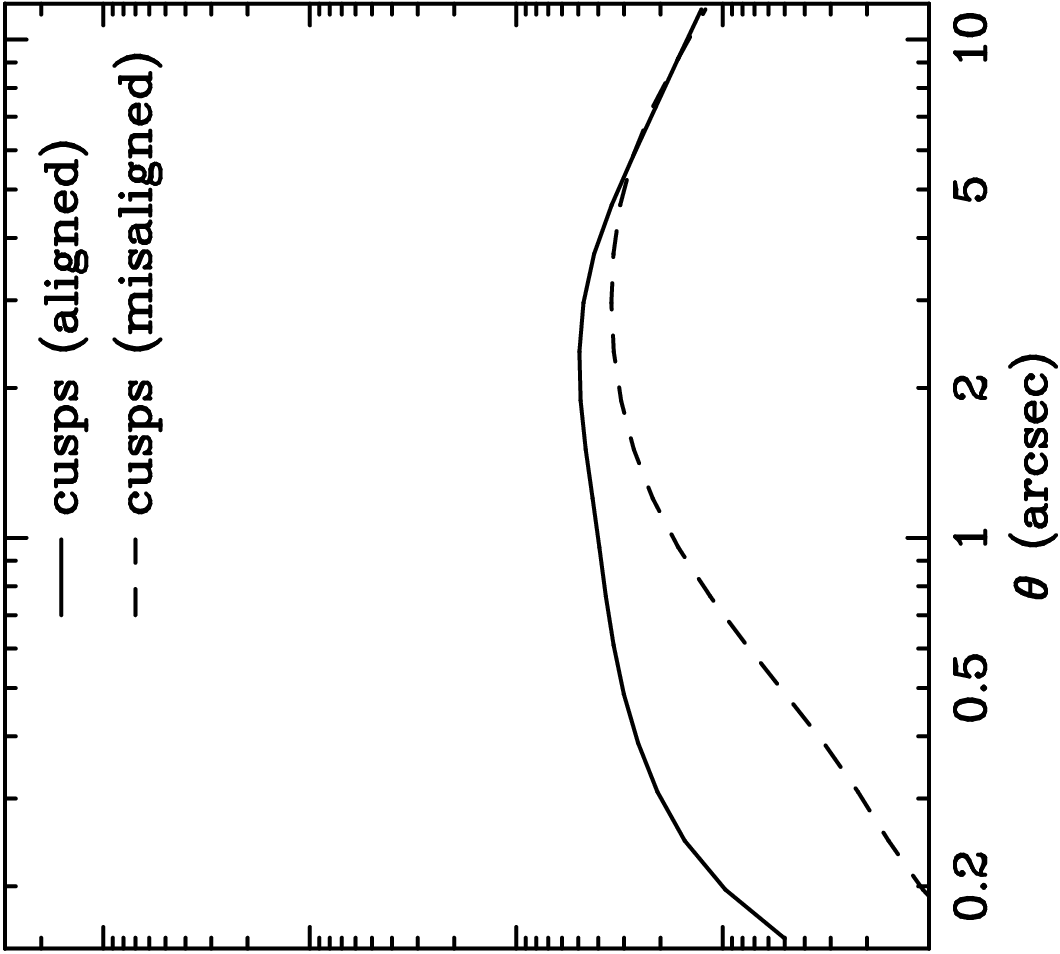}
		\label{fig:dpdtheta-0-90.cusps}
	}
	\subfigure[all]
	{
		\includegraphics[height=0.23\hsize,width=0.28\hsize,angle=-90]{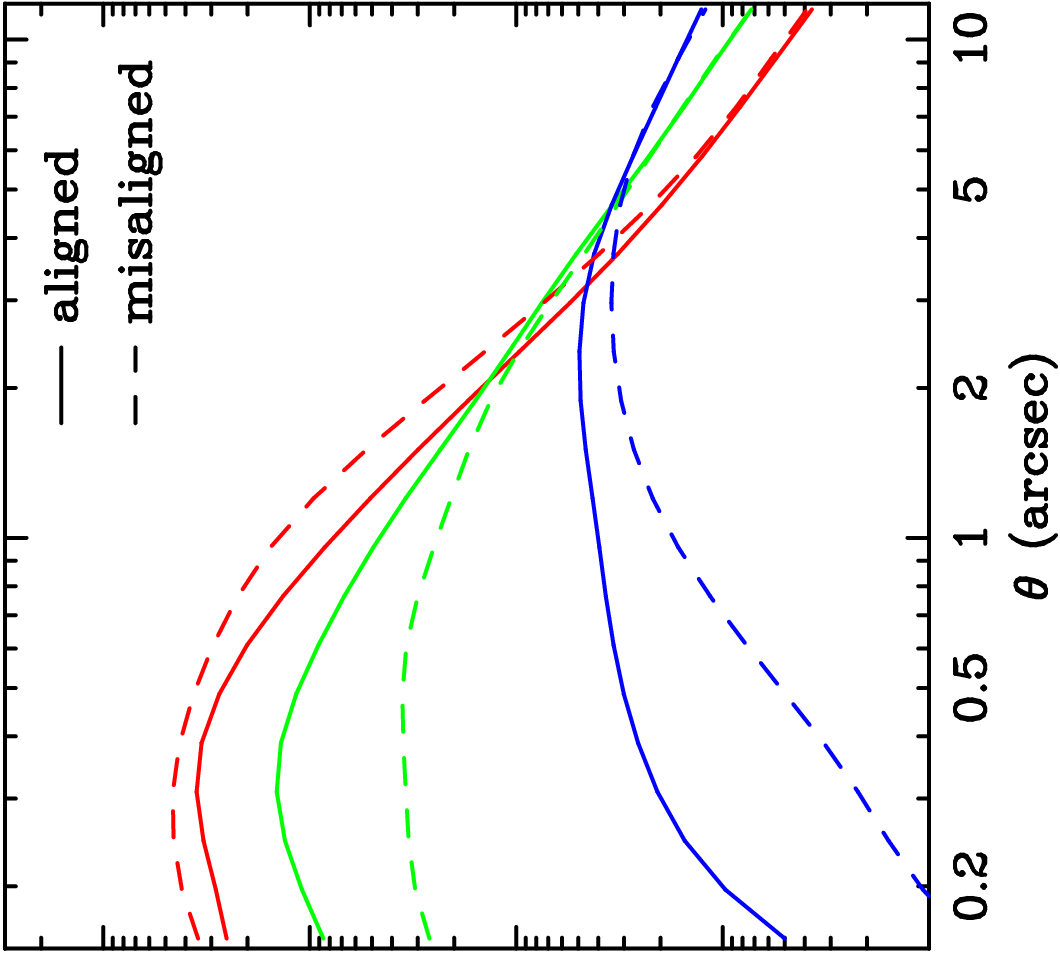}
		\label{fig:dpdtheta-0-90.all}
	}
	\caption{{\em Comparison of the distribution of image
            separations in the cases where the shapes of the luminous
            and dark matter projected densities are aligned and
            anti-aligned.} From left to right we show the
          distributions for doubles, quads and cusps. The last panel
          shows all three multiplicities in one plot. We have fixed
          the galaxy (luminous) axis ratio to be $q_H = 0.67$, and the
          ``second brightest image'' magnification bias is used here.}
	\label{fig:dpdtheta-0-90}
\end{figure*}

How much does the assumption of alignment vs. misalignment affect the
quad/double and cusp/double ratio? In fig.~\ref{fig:dpdtheta-0-90} we
plot these two cases. (Here, for simplicity we set the galaxy axis
ratio equal to $q_h = 0.67$, which is the mean axis ratio in the
``Coma model''.) The quad/double ratio at $\Theta = 0.3''$ changes
from ~$0.37$ in the aligned case, to ~$0.08$ in the unaligned
case. The cusp/double ratio changes from ~$6.5\times 10^{-3}$ in the
aligned case to ~$4.4\times 10^{-4}$, over a factor of 10 difference!
We must stress, however, that if galaxies and halos are randomly
aligned, the image multiplicities will fall in between the two extreme
cases we have plotted here. The total probability is very similar in
either case, changing from $4.8\times 10^{-4}$ in the aligned case to
$5.2\times 10^{-4}$ in the unaligned case (again at $\Theta =
0.3''$). The similarity is to be expected since only the overall shape
of the lensing mass is being changed.  The reason for the slight
enhancement in total probability for the unaligned case is due to the
magnification bias; this effect will be further explained later when
we compare to the lensing probability with spherical mass
distributions \S \ref{sec:results-spherical}.

\subsection{Effect of adiabatic contraction}

\begin{figure}
	
	\includegraphics[height=1.0\hsize,angle=-90]{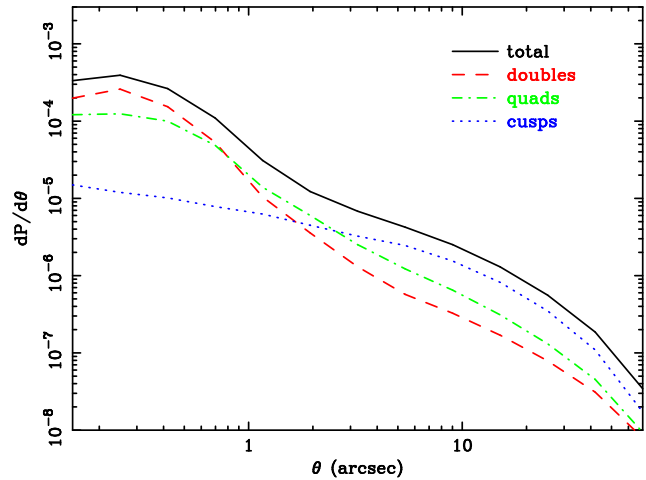}
\caption{{\em Distribution of image separations when no effect from
    baryon cooling is included.}  See also Fig. \ref{fig:dpdtheta-ac-noac}
  for a comparison of the cases with and without adiabatic
  contraction. The shapes of the projected luminous
  and dark matter densities are aligned and the axis ratio of the
  luminous component is fixed at $q_H = 0.67$.}
\label{fig:dpdtheta-no-ac}
\end{figure}

\begin{figure*}
	\centering
	\subfigure[total]
	{
		\includegraphics[height=0.265\hsize,width=0.28\hsize,angle=-90]{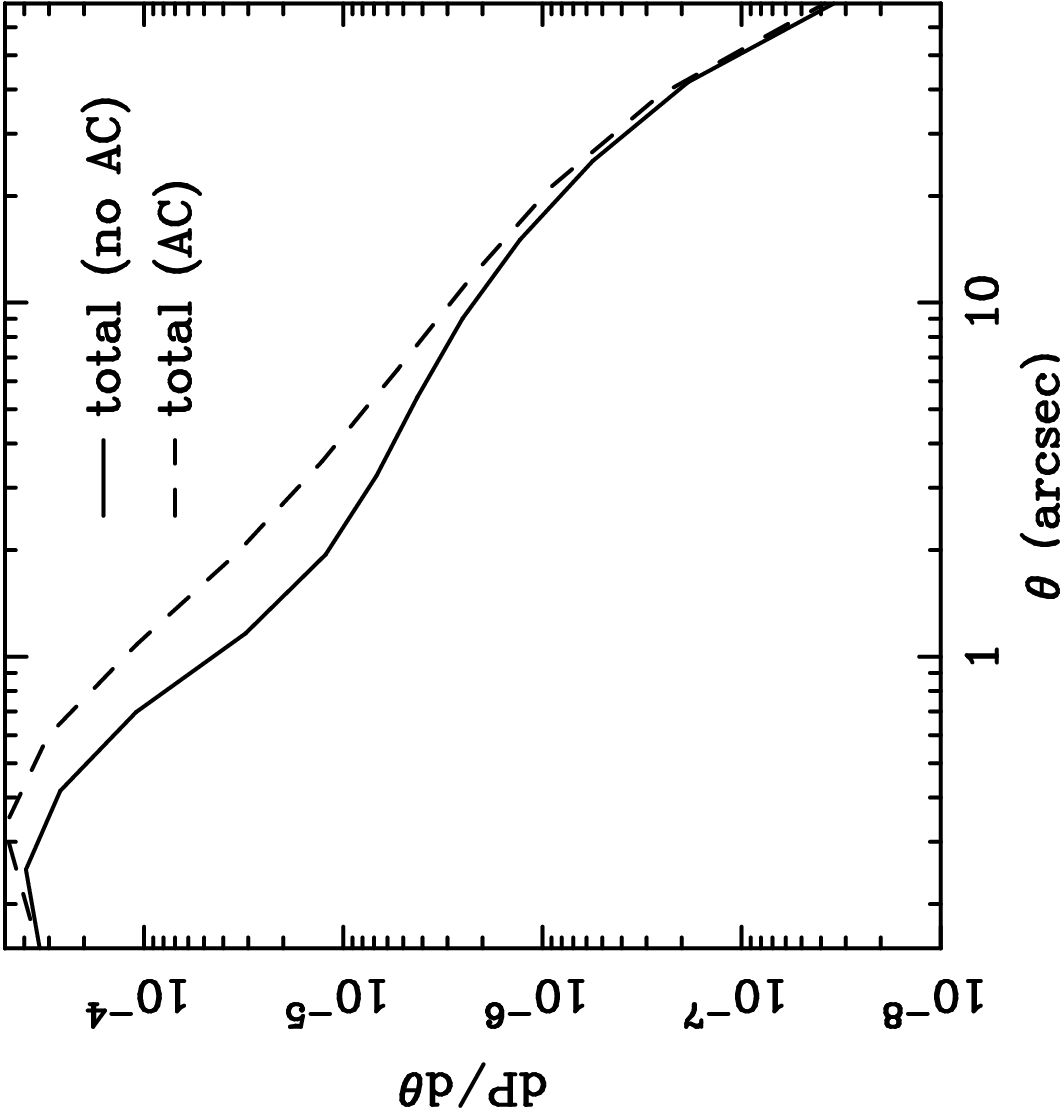}
		\label{fig:dpdtheta-ac-noac.total}
	}
	\subfigure[doubles]
	{
		\includegraphics[height=0.23\hsize,width=0.28\hsize,angle=-90]{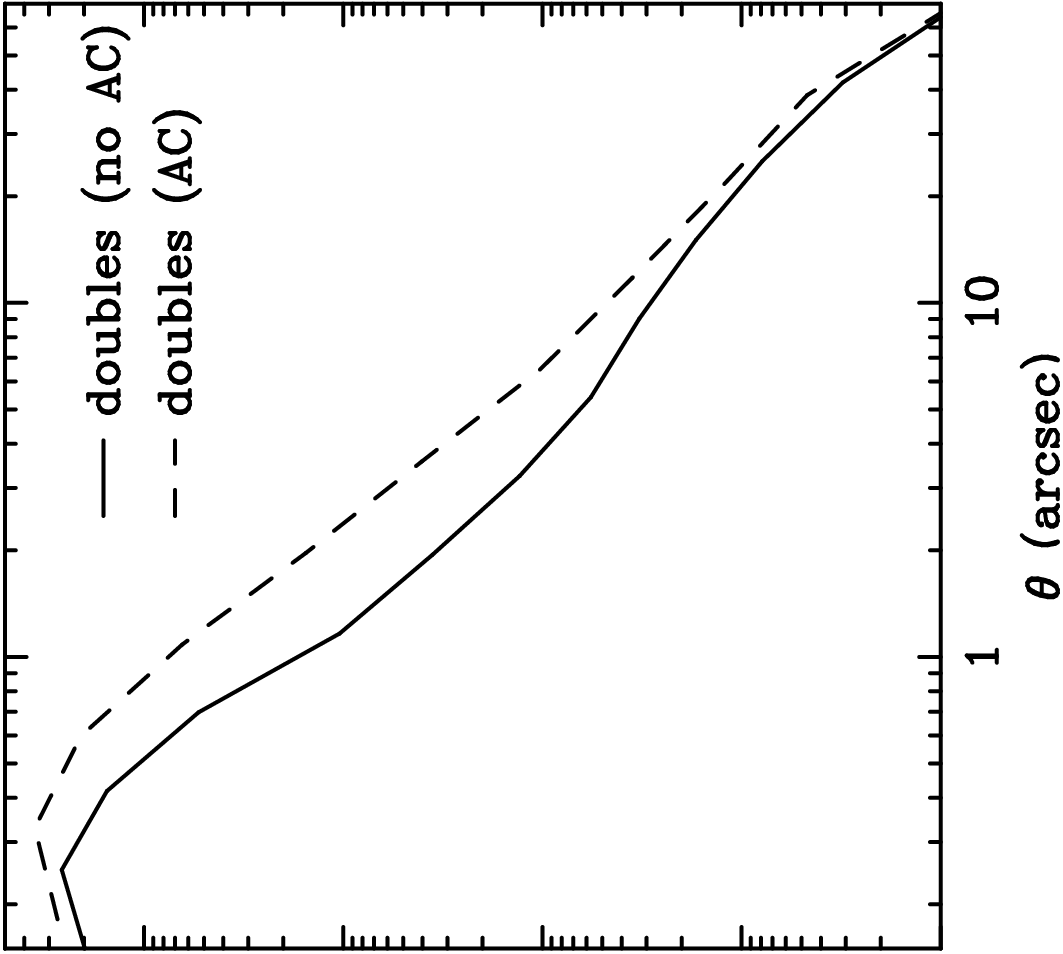}
		\label{fig:dpdtheta-ac-noac.doubles}
	}
	\subfigure[quads]
	{
		\includegraphics[height=0.23\hsize,width=0.28\hsize,angle=-90]{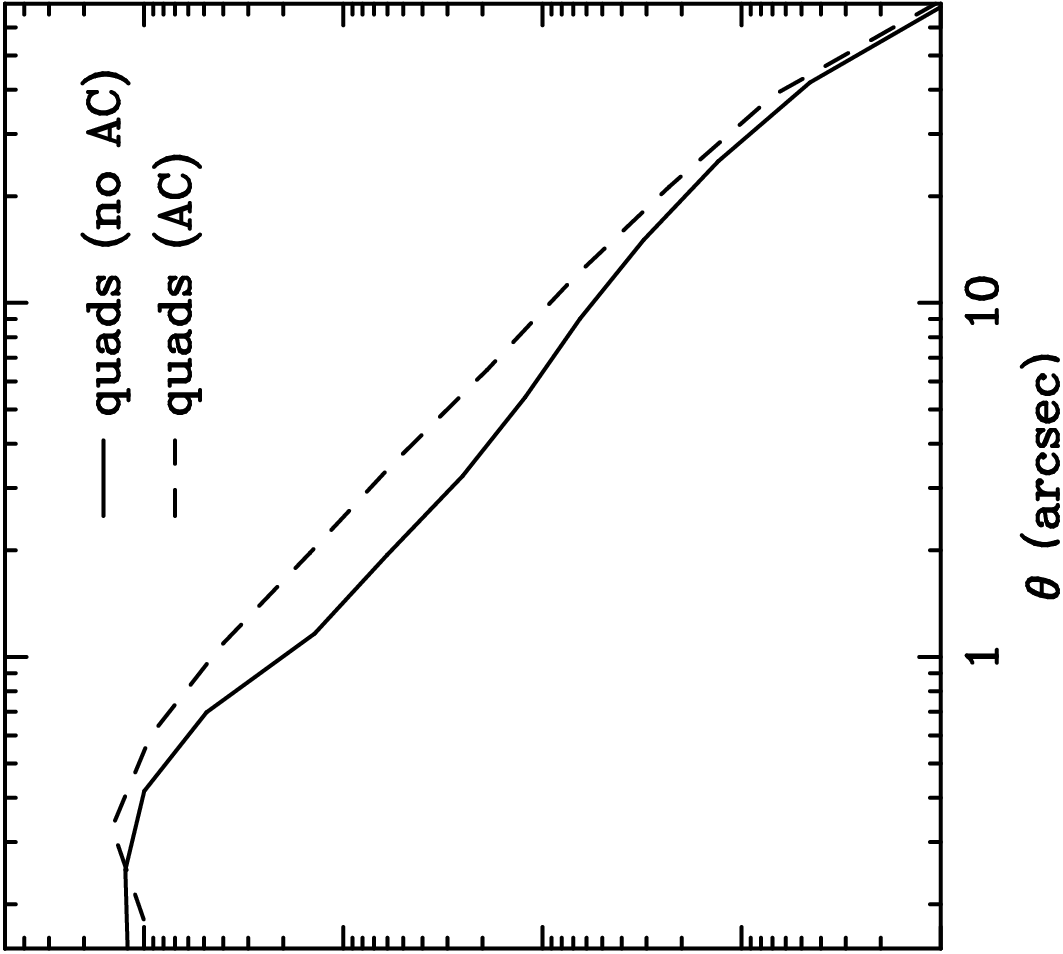}
		\label{fig:dpdtheta-ac-noac.quads}
	}
	\subfigure[cusps]
	{
		\includegraphics[height=0.23\hsize,width=0.28\hsize,angle=-90]{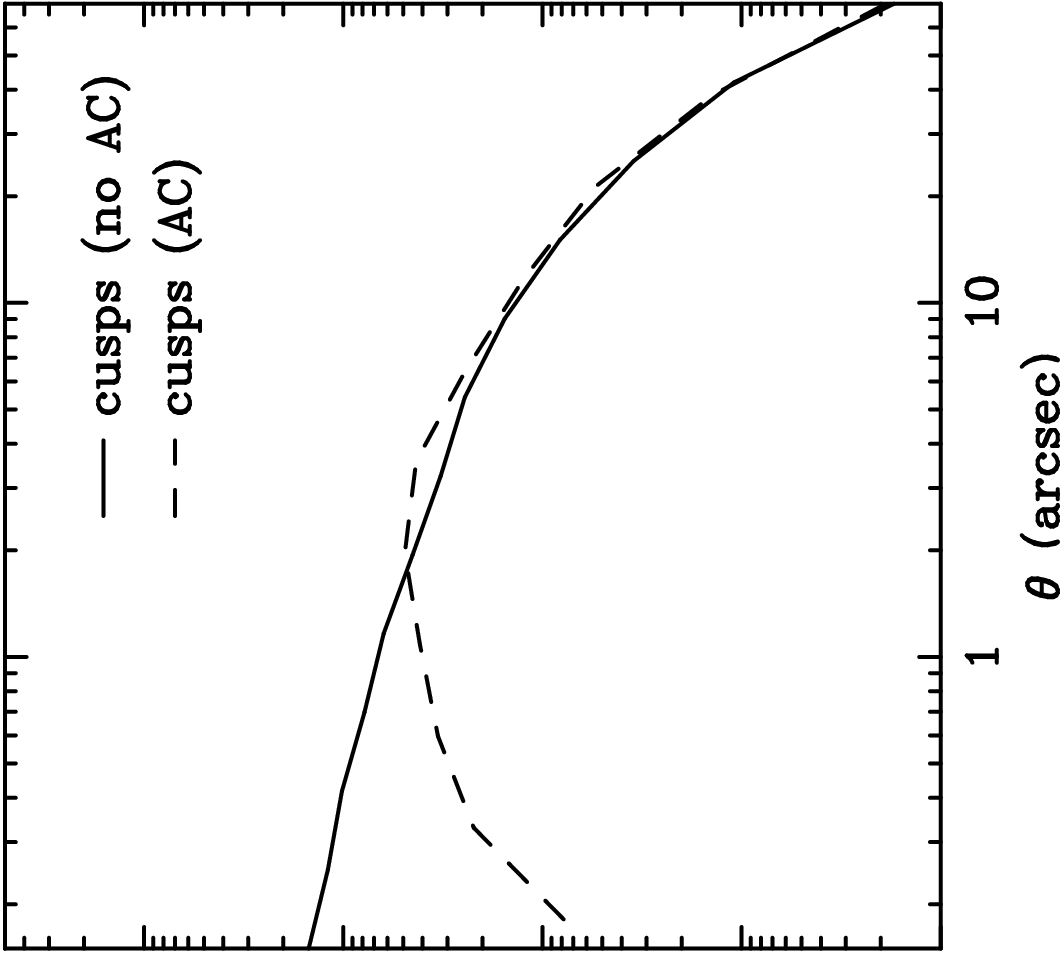}
		\label{fig:dpdtheta-ac-noac.cusps}
	}
	\caption{
{\em Comparison of the distribution of image separations when
  the effect from baryon cooling on dark matter halo (adiabatic
  contraction) is included (dotted) and not included (solid).}  The
shapes of the projected luminous  
  and dark matter densities are aligned and the axis ratio of the
  luminous component is fixed at $q_H = 0.67$. Going from left
        to right we have the comparisons for the the total lensing
        probability, doubles, quads, and cusps.} 
\label{fig:dpdtheta-ac-noac}
\end{figure*}

We now investigate what happens if adiabatic contraction is ``switched off'', 
i.e.  if the dark matter halo has an uncontracted triaxial NFW density
profile.   
The results are plotted in fig.~\ref{fig:dpdtheta-no-ac}. Here the galaxy and 
dark matter halo are aligned, and we have fixed the galaxy axis ratio at $q_H = 
0.67$.  The inner slope of the density profile is shallower compared to the AC 
case, resulting in more quads and cusps relative to doubles.  The point where 
quads and doubles become equal has shifted down to $\Theta = 0.8''$, so the 
predicted number of quads increases significantly compared to the adiabatically 
contracted model.  On the other hand, cusps become dominant at $2.5''$, which 
is inconsistent with the data since at most one naked cusp has been observed to 
date. For comparison, we plot the AC/no-AC cases together in 
fig.~\ref{fig:dpdtheta-ac-noac}. Not surprisingly, the lensing probabilities in 
the middle region are markedly higher in the AC case since the dark matter 
halos are more centrally concentrated than their original NFW form.

We also consider what happens if the axis ratios $a/c$, $b/c$ are increased by 
a factor of 0.1 following adiabatic contraction. The result is qualitatively 
the same as before, so we will not plot it here; however the turnover points 
are shifted toward higher image separations. Quads dominate over doubles at 
$3''$, so their overall fraction is predictably reduced. Cusps have shifted 
from ~$6''$ to ~$10''$, which may be more consistent with the data since, out 
of the few observed lenses with separations greater than $6''$, none of them 
have three observed images.  The quad fraction is obviously less consistent 
with the data in this case.  We expect however that numerous factors which 
increase the quad/cusp fraction, such as spiral galaxies and galaxy
environments,  
will have a greater effect at galaxy- and group-size separations than at 
cluster-size scales where cusps become dominant.  Thus, models where the 
overall axis ratios are increased by adiabatic contraction may ultimately be 
consistent with the lens data.

\subsection{Comparing to spherical mass
  distributions\label{sec:results-spherical}} 

\begin{figure}
	\includegraphics[height=1.0\hsize,angle=-90]{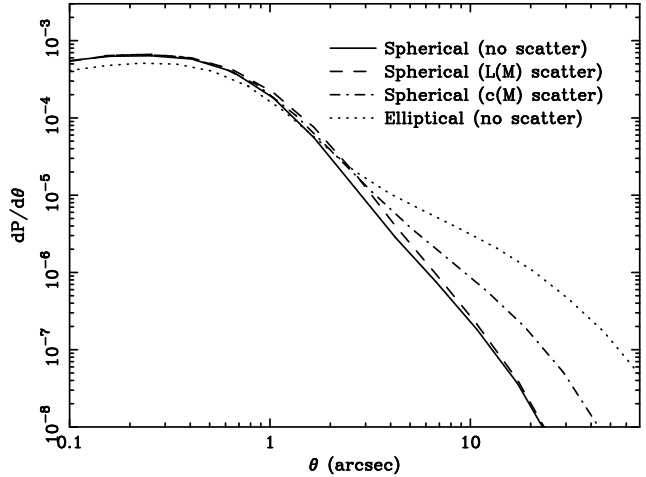}
\caption{{\em Comparison of the distribution of image separations in
    models with spherical and elliptical projected mass profiles.} 
For the spherical case, we separately also include scatter in the
mass-luminosity and concentration-mass relations. We
plot the total lensing probability in the triaxial case (same distribution 
as fig.~\ref{fig:dpdtheta-aligned-corr}) for comparison -- see curve
labeled ``elliptical''. The ``second-brightest image''  magnification
bias is used for all the calculations in this plot. } 
\label{fig:dpdtheta-spherical-comp}
\end{figure}

The difference between the spherical and elliptical case is striking
as may be ascertained from fig.~\ref{fig:dpdtheta-spherical-comp}.  The 
dramatic enhancement at large separations is a consequence of the triaxiality 
of the dark matter halos; elongation of halos along the line-of-sight creates 
considerably larger deflection angles (\cite{oguri-keeton04}). The enhancement 
gets larger with increasing separation because larger mass halos tend to be 
more triaxial.  The small-separation side does not show this effect because we 
did not start with a triaxial model for the central galaxy--rather, the galaxy 
properties were taken from observational data which naturally sees the galaxies 
in projection.

Curiously, the lens probability in the spherical case at small separations is 
actually higher than in the elliptical case. This is a consequence of the 
magnification bias. A source placed at the center of a lens with circular 
symmetry forms an Einstein ring; this degeneracy means that sources placed near 
the center will have both of their images highly magnified. A similar lens with 
ellipticity, however, does not have this degeneracy. Although sources placed 
inside and near the astroid caustic will be strongly magnified, a source 
slightly outside the astroid caustic does not produce such high magnifications 
since it maps to transition loci rather than critical curves (\cite{finch02}).  
This effect implies that if a lens with circular symmetry is deformed into an 
elliptical lens such that the average image separation stays fixed, the 
magnification bias for doubles is drastically reduced while the high 
magnification bias for quads may or may not compensate for this reduction.  
Whether the total bias cross section is increased or reduced by ellipticity 
depends on the type of bias used, and on the density profile of the lens. For 
our lens model, if the ``second brightest image'' bias is used (as in 
fig.~\ref{fig:dpdtheta-spherical-comp}) the total cross section will be lower 
than in the spherical case, whereas if the total magnification bias is used it 
will be roughly the same. Hence, depending on which type of bias is appropriate 
for a set of observed lens systems, the assumption of spherical galaxies may 
result in a slight overestimate of the total lensing probability.

\subsection{Scatter in mass-luminosity and concentration-mass
  relation}

In calculating all the results presented thus far, we ignored the effect of 
scatter in the mass-luminosity relation and in halo concentrations. The 
question naturally arises as to how scatter effects the lensing probabilities.  
To investigate this we calculate the total lensing probability assuming 
spherical galaxies and halos. The result is 
fig.~\ref{fig:dpdtheta-spherical-comp}, with the total lensing probability in 
the elliptical case (from fig.~\ref{fig:dpdtheta-aligned-corr}) being plotted 
for comparison. For the spherical concentrations of halos we used the 
\cite{bullock01} model, assigning log-normal scatter around the median with 
standard deviation $\sigma = 0.3$.  We also assumed log-normal scatter in 
mass-luminosity with a standard deviation $\Sigma = 0.25$, as derived in 
\cite{cooray05}.  Scatter in mass-luminosity has the effect of increasing the 
total lens probability by $\approx$11\%.  Scatter in concentration increases 
the contribution of the dark matter to the lensing probability, and has a 
noticeable effect at separations as small as $2''$.  While including scatter in 
these relations will not have a profound effect on image multiplicities, the 
concentration scatter will probably increase the quad fraction slightly in the 
middle region ($\approx 2''$-$10''$) since the dark matter halos produce higher 
quad and cusp fractions than the galaxies.

\section{Discussion}\label{sec:discussion}

We have shown that the distribution of strong lensing image multiplicities 
offers new possibilities for constraining correlations between galaxies and 
their host dark matter halos---in particular, adiabatic contraction and the 
degree of axial alignment. Since varying the amount of alignment makes a 
considerable difference in the quad/double ratio, and given that the observed 
ratio is so high, one might expect galaxies and halos to be quite closely 
aligned.  Simulations seem to support this (\cite{bailin05}, 
\cite{kazantzidis04}), and indeed there is some observational evidence from 
lensing: \cite{kochanek02} used mass profile modeling of 20 lenses from the 
CfA-Arizona Space Telescope Lens Survey (CASTLES) to show that the mass and 
light distributions were aligned within $\langle \Delta \theta^2\rangle^{1/2} < 
10^\circ$ where $\theta$ is the projected alignment angle.

The distribution of image multiplicities 
provides a simple method, far less time-consuming than individual lens 
modeling, that can be applied to a large statistical sample of
lenses. This method also has the advantange that the data for each
individual lens does not have to be very high quality. The
disadvantage is that we have to include all correlations and enviromental
effects that are either constrained by other data or directly obtained
from theory. 

One important caveat to keep in mind is that alignment of the
major/minor axes of galaxy and  halo does not necessarily imply that
the \emph{projected} axes will be aligned.   
The projected axes can also be misaligned if the galaxy and halo have differing 
triaxialities (\cite{keeton97}), analogous to the phenomenon of ``isophote 
twist''. Our method constrains the alignment of the projected axes. 
If the projected axes are shown to be closely aligned, then we
can draw the strong conclusion that the major/minor axes of galaxy and
halo are aligned, and that the triaxial galaxy and halo shapes are
similar on average.  

There are several ways in which our model can be improved. First, a fully 
triaxial model for adiabatic contraction from simulations is required. We used 
analytic fitting functions from the \cite{gnedin04} model of adiabatic 
contraction that was derived from halos with spherically averaged density 
profiles, extending it to triaxial halos by making the replacement $ r 
\rightarrow R$ where $R$ is the triaxial radius. But this is clearly only an 
approximation that may break down at high ellipticity.  The predicted lensing 
probability for cusps may differ significantly when a more accurate model of 
adiabatic contraction is used.

In addition, we have neglected the contribution from spiral galaxies.  This is 
partly justified by the fact that at least 75\% of galaxy lensing is thought to 
be due to ellipticals (\cite{moeller07}). However, since disk galaxies viewed 
edge-on can have extremely high projected ellipticities, their effect on image 
multiplicities may be substantial. We have also not included scatter in the 
mass-luminosity relation and in halo concentration, the effects of which were 
discussed in the previous section. Although the image multiplicities are not 
drastically effected by the scatter, it is nevertheless an important effect and 
has an appreciable impact on the total lensing probability.

We have also ignored the effect of external shear from lens galaxy 
environments, which can increase lensing probabilities at intermediate image 
separations, $> 3''$ (\cite{oguri05}).  Another effect is the misalignment of 
the center-of-mass of galaxies with that of their host halos, which will 
slightly enhance the cross section for quads (\cite{quadri03}).  Perhaps most 
importantly, we have not included substructure in our calculations.  Work by 
\cite{cohn04} suggest that satellites can roughly double the expected 
quads-to-doubles ratio, thus dramatically altering the predicted ratio from our 
model. They also give rise to cross sections for more exotic image 
configurations, with five or more visible images, whose lens probabilities may 
be significant.  Including all these effects will be essential for deducing 
rigorous constraints on the profile, shape and correlations of lensing
masses from future data.

\section*{Acknowledgements}
We have extensively used Charles Keeton's \emph{Gravlens} code during
this work to test our code. We thank Charles Keeton for making his
lensing software public. 

\bibliography{lens}
\bibliographystyle{mn2e}

\label{lastpage}

\end{document}